\documentclass[aps,prb,reprint]{revtex4-1}
\usepackage{graphicx,color}
\usepackage{epstopdf}
\usepackage{color}

\begin{document}
\title{Nuclear magnetic resonance study of thin Co$_2$FeAl$_{0.5}$Si$_{0.5}$ Heusler films with varying thickness}

\author{A.~Alfonsov$^1$\footnote{Electronic address: a.alfonsov@ifw-dresden.de}, B.~Peters$^2$, F.Y.~Yang$^2$, B.~B\"uchner$^{1,3}$, S.~Wurmehl$^{1,3}$\footnote{Electronic address: s.wurmehl@ifw-dresden.de}}

\affiliation{$^1$Leibniz Institute for Solid State and Materials
Research IFW,
Institute for Solid State Research, 01069 Dresden, Germany\\
$^2$Department of Physics, The Ohio State University, Columbus, OH
43210\\
$^3$Institute for solid state physics, Technische Universit\"at
Dresden, D-01062 Dresden, Germany}


\date{\today}

\begin{abstract}
Type, degree and evolution of structural order are important aspects for understanding and controlling the properties of highly spin polarized Heusler compounds, in particular with respect to the optimal film growth procedure. In this work, we compare the structural order and the local magnetic properties revealed by nuclear magnetic resonance (NMR) spectroscopy with the macroscopic properties of thin Co$_2$FeAl$_{0.5}$Si$_{0.5}$ Heusler films with varying thickness. A detailed analysis of the measured NMR spectra presented in this paper enables us to find a very high degree of $L2_1$ type ordering up to 81\% concomitantly with excess Fe of 8 to 13\% at the expense of Al and Si. We show, that the formation of certain types of order do not only depend on the thermodynamic phase diagrams as in bulk samples, but that the kinetic control may contribute to the phase formation in thin films. It is an exciting finding that Co$_2$FeAl$_{0.5}$Si$_{0.5}$  can form an almost ideal $L2_1$ structure in films though with a considerable amount of Fe-Al/Si off-stoichiometry. Moreover, the very good quality of the films as demonstrated by our NMR study suggests that the novel technique of off-axis sputtering technique used to grow the  films sets stage for the optimized performance of Co$_2$FeAl$_{0.5}$Si$_{0.5}$ in spintronic devices.
\end{abstract}

\maketitle

\section{Introduction}

Spintronics is considered a potential follow-up technology to
purely charge-based electronics. In spintronic devices, both
charge and spin of electrons are used as information carriers
leading to faster switching at lower energy consumption compared
to charge-based electronics. Half-metallic ferromagnets (HMFs)
are the optimal materials to be implemented in spintronic devices
\cite{GME83,CVB02,FFB07,IIT08,GFP11} as their conduction electrons
are expected to be 100\% spin polarized. Heusler compounds with
$L2_1$ type structure represent an especially favorable family of
predicted HMF compounds and seem to offer all necessary
ingredients for their implementation in spintronic devices such as
high spin polarization,\cite{GME83,FFB07,GFP11,BBV13} high Curie
temperatures,\cite{WFK05b,WFK06} and a low Gilbert damping
constant.\cite{TTO09} However, the observation of the required
key spintronic properties in Heusler compounds crucially depends
on the type and degree of structural ordering.\cite{IIT08,IWJ08,GFP11}

Nuclear magnetic resonance spectroscopy (NMR) allows one to probe
the local environments of $^{59}$Co nuclei in Co-based Heusler
bulk and film samples, and, thus, enables characterization of local
order and quantification of different structural contributions concomitantly with an off-stoichiometric composition.\cite{pan97,IOM06,IWJ08,WK08,WKS07,WKS08,WKS09,WJK11,RAB13} Such a local probe of
structure and composition is very useful since compounds comprising
elements from the same periodic row (e.g.,\ Co and Fe) have very
similar scattering factors for x-rays, and thus, x-ray diffraction
(XRD) only may not be sufficient to resolve the structural
ordering unambiguously in particular if both disorder and deviations from the 2:1:1 stoichiometry are present.\cite{BWF07}

In addition to information on the chemical, crystallographic
environments, the NMR technique is useful to determine the magnetic
state of a ferro- or ferrimagnetic material. The restoring field
($H_{rest}$) is an effective magnetic field originating from a
resistance to magnetic oscillations, and therefore is proportional
to the square root of the optimal power (i.e.\ the power producing
the maximum spin-echo intensity) of the applied rf pulses during
an NMR experiment. $H_{rest}$ derived in NMR experiments provides
a measure of magnetic stiffness or magnetic anisotropy on a local
scale, compared with the macroscopic domain wall stiffness
contributing to the coercive fields from SQUID magnetometry.\cite{pan97,WK08} The advantage of NMR is that we can measure at
a given frequency and can thus relate the magnetic stiffness to a
specific local magnetic environment (e.g.,\ phase or structure).

A particular interesting Heusler compound to be mentioned in the
context of HMF is Co$_2$FeAl$_{0.5}$Si$_{0.5}$. Band structure
calculations predict a high stability of the minority band gap in
this compound,\cite{FFB07,BFF07} a prediction which is  experimentally supported.\cite{WJS12} Co$_2$FeAl$_{0.5}$Si$_{0.5}$ has
been implemented in thin films and magnetic tunnel junctions.\cite{TIM06,TIS06,TIM07,TIS07b,TIM09,NRG07,SSW09,TMS12,Tez12}
Recently, we have epitaxially grown Co$_2$FeAl$_{0.5}$Si$_{0.5}$
films on lattice-matched MgAl$_2$O$_4$ (001) substrates by a novel
off-axis sputtering technique yielding films with an exceptionally
high quality.\cite{PAB13} In this work, we characterize the local
crystallographic and magnetic structure of these films using NMR. We were able to relate the
macroscopic physical properties of these
Co$_2$FeAl$_{0.5}$Si$_{0.5}$ films to the local ordering.

\section{Experimental Details}

Epitaxial Co$_2$FeAl$_{0.5}$Si$_{0.5}$ films were grown on
MgAl$_2$O$_4$ (001) substrates by off-axis sputtering in a UHV
system with a base pressure as low as 7$\times$10$^{-11}$~Torr
using ultra-pure Ar (99.9999\%) as sputtering gas. Optimal quality
Co$_2$FeAl$_{0.5}$Si$_{0.5}$ epitaxial films were obtained at an
Ar pressure of 4.5~mTorr, a substrate temperature of 600$^\circ$~C,
and DC sputtering at a constant current of 12~mA, which results in
a deposition rate of 5.6~${\rm\AA}$/min.
The Co$_2$FeAl$_{0.5}$Si$_{0.5}$ epitaxial films were characterized
by a Bruker D8 Discover high-resolution triple-axis x-ray diffractometer (XRD).
Details about growth and characterization are found elsewhere.\cite{PAB13}

The NMR experiments were performed at 5~K in an automated,
coherent, phase sensitive, and frequency-tuned spin-echo
spectrometer ({\it{NMR Service}} Erfurt, Germany). We used a
manganin coil wrapped around the sample to apply and pick up the
rf pulses. This coil is implemented in an LC circuit with three
capacitors. The NMR spectra were recorded at 5~K in the frequency
($\nu$) range from 104-254~MHz in steps of 0.5~MHz in zero magnetic
field. All NMR spectra shown here were corrected for the
enhancement factor as well as for the $\nu^2$ dependence,
resulting in relative spin-echo intensities which are proportional
to the number of nuclei with a given NMR resonance frequency.\cite{pan97,WK08}

\section{Results and discussion}

\begin{figure}
\centering
\includegraphics[width=8cm]{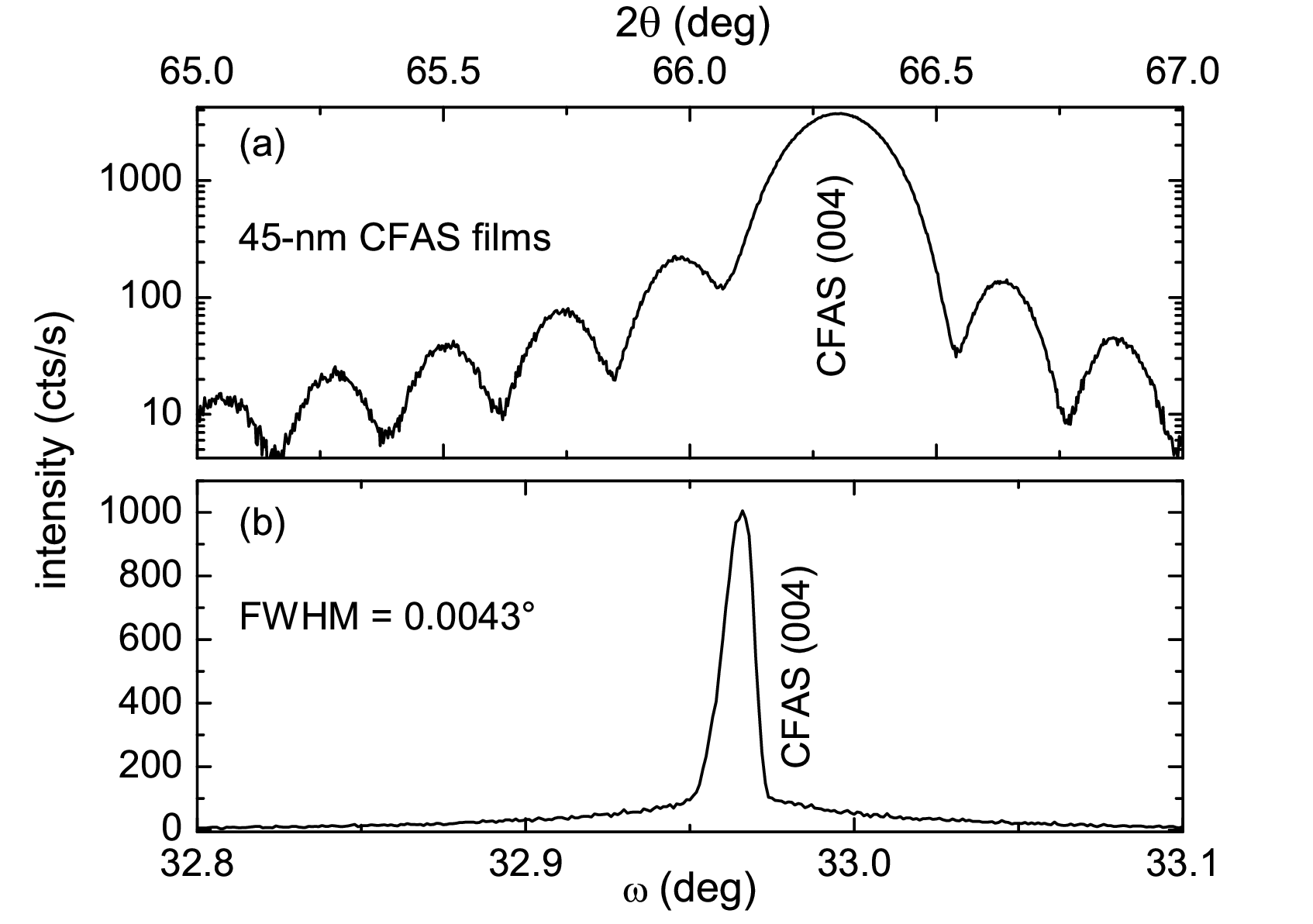}
\caption{(a) High resolution $\theta/2\theta$ XRD scan of a 45-nm Co$_2$FeAl$_{0.5}$Si$_{0.5}$ (CFAS) film grown on MgAl$_2$O$_4$ (001) substrates. (b) XRD rocking curve of the (004) peaks of the Co$_2$FeAl$_{0.5}$Si$_{0.5}$ film gives a FWHM of 0.0043$^\circ$.}
\label{Fig_xrd}
\end{figure}

Figure~\ref{Fig_xrd} shows the $\theta/2\theta$ scan of a 45-nm thick Co$_2$FeAl$_{0.5}$Si$_{0.5}$ epitaxial film on MgAl$_2$O$_4$ (001). The clear Laue oscillations near the Co$_2$FeAl$_{0.5}$Si$_{0.5}$ (004) peak demonstrate the high crystalline uniformity as well as smooth surface and sharp interface with the substrate. Fig.~\ref{Fig_xrd}(b) presents a rocking curve of the (400) peak with a full-width-at-half-maximum (FWHM) of 0.0043$^\circ$, which is at the instrumental resolution limit of our high-resolution XRD system, revealing exceptional crystalline quality.

\begin{figure}
\centering
\includegraphics[width=8cm]{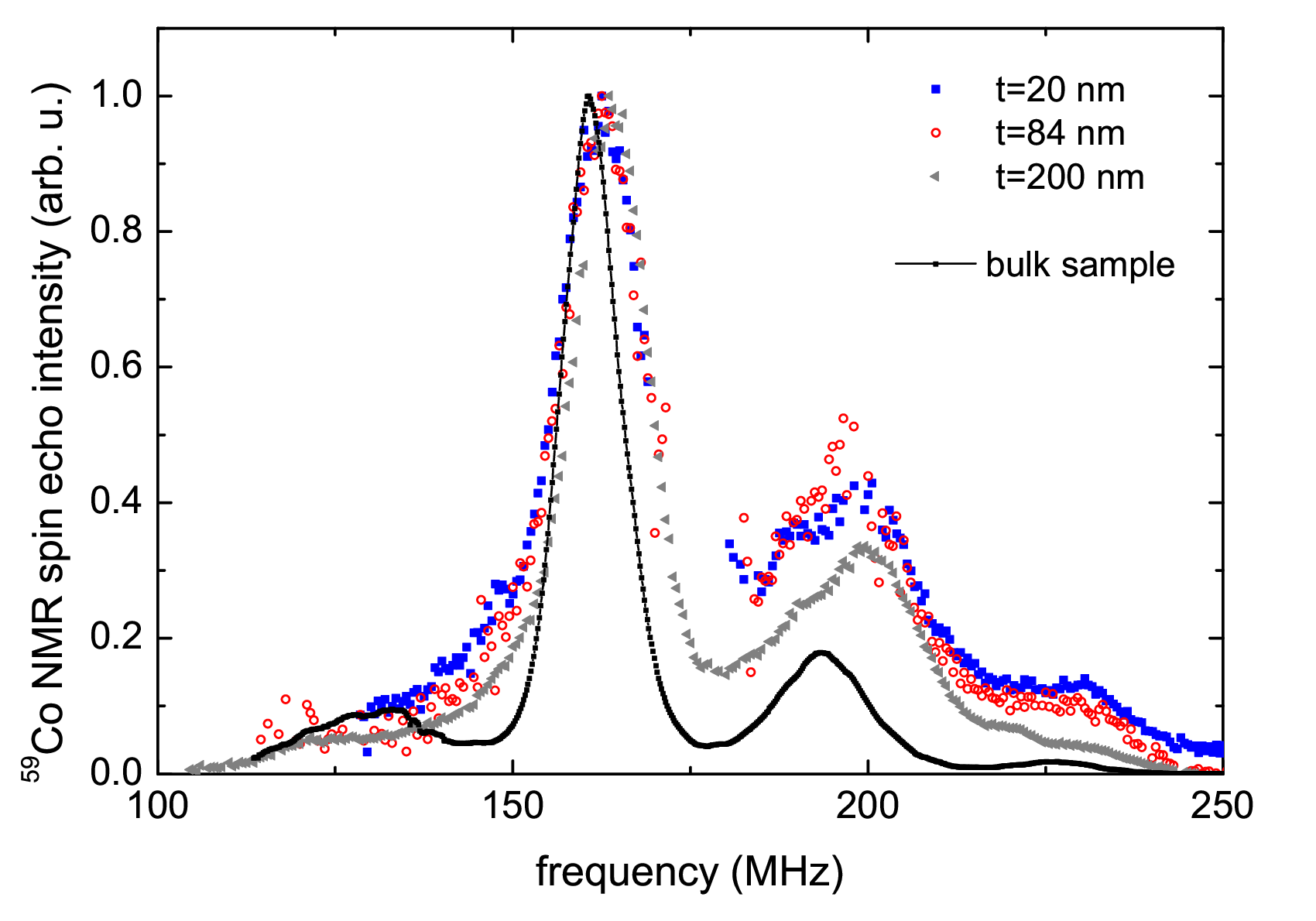}
\caption{(Color online) Normalized $^{59}$Co NMR spectra of Co$_2$FeAl$_{0.5}$Si$_{0.5}$ thin films with thicknesses of $t$ = 20, 84, and 200~nm in comparison with the NMR data of a Co$_2$FeAl$_{0.5}$Si$_{0.5}$ bulk sample.\cite{WKS12} Note that missing data points in the middle of the spectra are due to the increased spectrometer noise for this frequency. }
\label{Fig_1}
\end{figure}

In order to further characterize the structural quality of our films we measured the $^{59}$Co NMR spectra for different thin film
samples with varying thickness (20, 45, 84, 120 and 200~nm).
Figure~\ref{Fig_1} exemplarily shows the normalized $^{59}$Co NMR
spectra of films with thickness t = 20, 84 and 200~nm in comparison
with that of a Co$_2$FeAl$_{0.5}$Si$_{0.5}$ bulk sample (data
taken from Ref.~\onlinecite{WKS12}). All spectra share the main line
around 163~MHz with one shoulder on the low frequency side and two
pronounced satellites on the high frequency side with spacing of about $\sim 33$~MHz between them.

The observation of low and high frequency satellite lines with
a spacing of about 33~MHz suggests a contribution from $B2$ type
ordering of the films, in line with the interpretation of the NMR
data for Co$_2$FeAl$_{0.5}$Si$_{0.5}$ and Co$_2$FeAl bulk samples.\cite{WKS12,WKS08} Partial $B2$ type
ordering of the films is consistent with the ternary thermodynamic Co-Fe-Al phase diagram at 650$^\circ$~C\cite{Koz99} and with experimental results for the Co$_2$FeAl$_{0.5}$Si$_{0.5}$ compound from Umetsu \textit{et al.}\cite{UOK12} Neglecting other contributions to the high frequency satellite, its higher intensity in the films may be understood in terms of higher degree of $B2$ type contributions. This interpretation, however, is in strong contrast to the significantly smaller NMR echo intensity at $\sim130~$MHz in films compared to the bulk sample. In fact, NMR spectra of the film samples do not exhibit a clear satellite, but rather a shoulder, on the low frequency side, which complicates the qualitative comparison of the spectra. Taking into account both observations, the larger high frequency satellites and the poorly resolved low frequency satellite in the films compared to the bulk sample, it seems natural to interpret this observation as a deviation from the 2:1:0.5:0.5 stoichiometry, and more specifically, to assume, that the films are more Fe-rich and Al/Si-poor than the expected 2:1:0.5:0.5 stoichiometry
(compare Ref.~\onlinecite{IWJ08,WKS09,WKS12}). The formation of Fe-rich
environments may also be responsible for the slightly higher than expected
magnetic saturation-moment as, according to the Slater-Pauling rule, the expected value for magnetic moment in case of Co$_2$FeAl$_{0.5}$Si$_{0.5}$ compound is $5.5~\mu$B/f.u., whereas the measured one for the 45~nm film is about $5.6~\mu$B/f.u. (see Ref.~\onlinecite{PAB13}).

Thereby, already a qualitative analysis of the thin films NMR spectra suggests a contribution from both $L2_1$ and $B2$ types of order, as well as a presence of an Fe to Al/Si off-stoichiometry. Kozakai \textit{et al.} \cite{Koz99} report that $B2$ is the thermodynamic stable phase at 600$^\circ$~C in Co$_2$FeAl. Whereas Umetsu \textit{et al.} report that the $L2_1$ type phase is thermodynamically stable below 1125~K and Co$_2$FeAl$_{0.5}$Si$_{0.5}$ undergoes a transition to $B2$ type order at 1125~K.\cite{UOK12} Please note, that even for the bulk sample annealed below the ordering temperature no full $L2_1$ order is realized.\cite{WKS12,UOK12} In the present case of a thin film, both $B2$ and the higher ordered $L2_1$ type phases are found, suggesting additional influence of e.g. the substrate, strain and/or kinetic contributions upon cooling.

In order to perform a detailed quantitative
analysis of all the contributions to the NMR spectra, we
fitted the NMR spectra of all samples using a sum of Gaussian
lines. The corresponding parameters of these lines, such as
resonance frequency, linewidth, and intensity, were
constrained according to a model similar to the one described in
detail in Ref.~\onlinecite{WKS12}. 

For $L2_1$ type order only one NMR line is expected, while $B2$
type order yields several NMR lines.\cite{WK08,WKS08} Hence, in
the presence of both $L2_1$ and $B2$ type order and
off-stoichiometry, the relative area of the NMR spectra can be
represented as a sum of several lines originating in different
structural and compositional contributions. The spacing between
adjacent resonance lines, $\Delta$$B2$, may be assumed to be a
constant while their relative contribution to the NMR spectrum is
given by the amount of random mixing of Fe and Al/Si on one
crystallographic site ($B2$ type structure) as well as by the Fe to Al/Si
ratio. The off-stoichiometry between Fe and Al/Si contributes to
NMR lines on the high frequency side only due to the extra Fe at
the Al/Si sites in the first Co shell. From the relative areas of
these lines, the amount of off-stoichiometry and $L2_1$/$B2$ type
order in the films can be quantified. Due to the random mixing of
Al and Si on one ($L2_1$ plus off-stoichiometry) or two ($B2$ plus
off-stoichiometry) crystallographic sites, each NMR line further
broadens or splits into a set of sub-lines with equal spacing
$\Delta$$\frac{Al}{Si}$ between them. This splitting originates in the small difference in the hyperfine field seen by Co nuclei depending on which atom, either Al or Si is located in the first coordination shell.\cite{BFF07,WKS12} Compounds with Si have one extra valence electron with respect to the compounds with Al. This extra valence electron increases the magnetic moment of the compound which in turn changes the transferred contributions to the hyperfine field and concomitantly the resonance frequency.\cite{BFF07,UOK12,WJS12} Hence, each specific configuration with particular Al and Si neighbors in the first shell of Co arising from the random distribution expected for a quaternary compound will have a different resonance frequency. For details see Supplemental Material at the page 8 and Ref.~\onlinecite{WJS12,WKS12}. The relative contributions of the Gaussian lines in the fit can be compared to the probabilities calculated from a random atom model,\cite{WKS12} which is mathematically expressed in form of a
binomial distribution function:

\begin{eqnarray}
\label{eq1}
   && P(n,m,l,k,x,u,y,C_{B2},C_{L2_1}) = \\ \nonumber
   && C_{B2} \frac{N!}{n!(N-n-m)!m!} \\ \nonumber
   && (1-x)^{N-(m+k)}  x^{m+k} y^k (1-y)^{(N-n)-k} +\\ \nonumber
   && C_{L2_1} \cdot \frac{L!}{l!(L-l-k)!k!} (1-u)^{L-l}  u^l y^{k}(1-y)^{(L-l)-k} \delta_{n,4} \nonumber
\end{eqnarray}
\\
\[ \text {with}~\delta_{n,4} \left \{\begin{array}{ccc}
  1, & \text{if }n\text{~ $\ge$ 4}\\
  0, & \text{if }n\text{~ $<$ 4}
\end{array}  \right.\]


The first term in Eq.\ref{eq1} represents the $B2$ contributions
with a random distribution of Fe and Al/Si, where $C_{B2}$ is the
degree of $B2$ type order. This random distribution involves both
the 4$a$ and 4$b$ Wyckoff positions of the respective $L2_1$
lattice, which correspond to the 1$b$ position in the $B2$
notation. In Eq.\ref{eq1}, $x$ represents the Fe to (Al+Si)
stoichiometry, enabling us to calculate the probability of
finding Fe atoms on the $Z$ (Al and Si) sites, and, hence, to quantify the Fe-Al,Si off-stoichiometry; $y$ denotes the Al
to Si stoichiometry (Al$_y$Si$_{1-y}$). For stoichiometric
Co$_2$FeAl$_{0.5}$Si$_{0.5}$ films with complete $B2$ type order,
$x$ = 0.5 while $x$ $>$ 0.5 indicates off-stoichiometry with Fe
excess. Here, $N = 8$ is the number of possible sites for atoms in
the first Co shell while $n$, $m$, and $k$ are the corresponding
numbers of Fe, Si, and Al atoms, respectively, in the first Co
shell (note $n+m+k=N)$. 

The second term in Eq.\ref{eq1} represents the
contribution from $L2_1$ type order, $u$ is the amount of Fe to
(Al+Si) off-stoichiometry ($u$ = 0 for stoichiometric
composition), $L=4$ is the number of possible sites for Fe
atoms on the Al/Si sites in the first Co shell , $l$ and $k$ are
the numbers of Fe and Si/Al atoms in the first shell,
respectively. Since both $x$ and $u$ represent the
off-stoichiometry, there is a relation between these two
parameters $x$ = 0.5 $u$ + 0.5. The coefficients $C_{B2}$ and
$C_{L2_1}$ represent the relative contribution from $^{59}$Co
nuclei with a $B2$ and $L2_1$ first shell environment,
respectively, and $C_{B2}$ + $C_{L2_1}$ = 1. There are two ways to realize the presence of both $L2_1$ and $B2$ in a given sample: Case (i) deals with large $B2$ type domains within a $L2_1$ matrix, where the number of Co nuclei located at the interfaces between both phases is negligible compared to the number of Co nuclei within a certain phase region in line with the recent report on Co$_2$MnSi films by T. Miyajima \textit{et al.}\cite{MOK09} In that case, the coefficients obtained from our binomial model immediately give the ratio between $L2_1$ and $B2$ phases. 
In the second case (ii) both $L2_1$ and $B2$ phases are so finely dispersed, that the number of Co nuclei at the interfaces is not negligible anymore. In that case, the Co nuclei at the interface experience surrounding similar to that of B2, and therefore the overall degree of order is even underestimated by our model. Moreover in this case the distribution is no longer described by Eq.\ref{eq1} (see Supplemental Material at the page 8 for details). Since our binomial model (Eq.\ref{eq1}) well fits to the measured NMR data (see below), scenario (i) seems to be valid in the present case, which is also in line with the recent report on Co$_2$MnSi films by T. Miyajima \textit{et al.}\cite{MOK09}

\begin{figure}
\centering
\includegraphics[width=8cm]{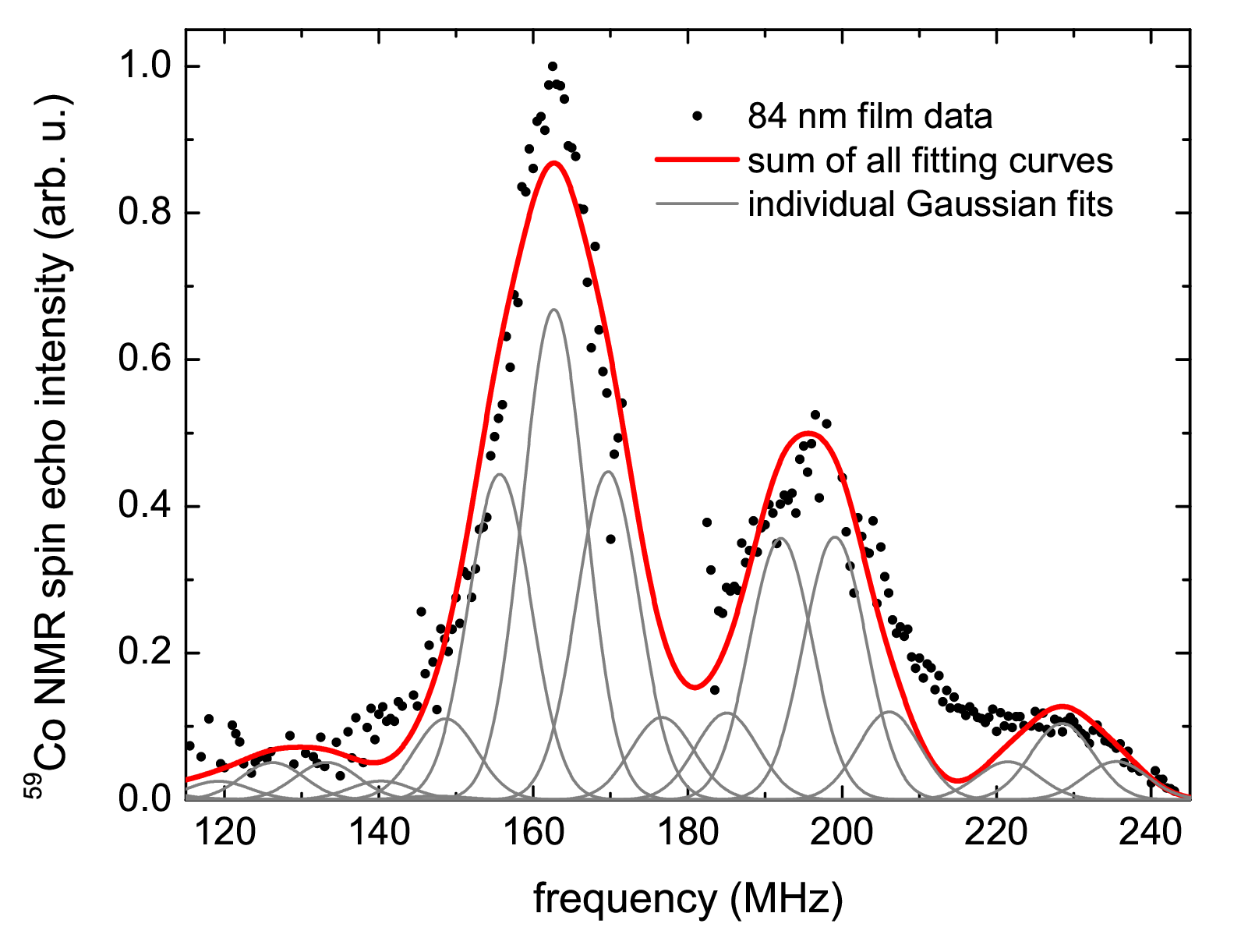}
\caption{(Color online) Normalized $^{59}$Co NMR spectrum of an 84-nm thick Co$_2$FeAl$_{0.5}$Si$_{0.5}$ film, shown together with a fitted curve (solid line). Analysis of the data gives a degree of $L2_1$ order of 81\%. }
\label{Fig_3}
\end{figure}

Figure~\ref{Fig_3} exemplarily shows the fitting result of the NMR spectrum
for the 84~nm Co$_2$FeAl$_{0.5}$Si$_{0.5}$ films where the respective fitting parameters are NMR resonance frequencies, the spacing between lines $\Delta$$B2$, $y$, $u$, $C_{B2}$, $\Delta$Al/Si, and the linewidths of individual Gaussian lines. The residual fit mismatch for all spectra
does not exceed 15\% which is quite good for such a rather
simple model. The fit yields the average spacing between the main line and the high frequency satellites of 33~MHz, which is slightly larger than in the corresponding bulk samples (31~MHz).\cite{WKS12} The spacing $\Delta$Al/Si between lines due to the mixing of Al and Si is found to be about 7~MHz, which is very similar to the bulk sample. In addition, the fit yields the Al to Si ratio of $0.5(\pm0.01):0.5(\pm0.01)$ as expected from the nominal composition. In general, Al and Si may not be homogeneously distributed in the Co$_2$FeAl$_{1-x}$Si$_x$ series. We have seen such an inhomogeneous distribution by NMR in the Co$_2$Mn$_{1-x}$Fe$_x$Si series, where Fe in the the Fe-rich samples is not entirely randomly distributed. Obviously, such a preferential order will not follow the random atom model as described in Eq.\ref{eq1} (also see Ref.~\onlinecite{WAK13}). Such an inhomogeneity is likely not present in the Co$_2$FeAl$_{0.5}$Si$_{0.5}$ films for two reasons: (i) for the corresponding bulk system Al and Si are fairly homogeneously distributed,\cite{WKS12} and (ii), we found no deviations from our random atom model hinting on such an inhomogeneous distribution of Al and Si, with the exception that the relatively large mismatch of the fit and the measured data at frequencies near $\sim140$~MHz and $\sim215$~MHz may be related to the additional contributions from Co$_2$FeSi and CoAl or fcc Co impurities, as suggested in Ref.~\onlinecite{WKS12} for the bulk sample.

\begin{figure}
\centering
\includegraphics[width=8cm]{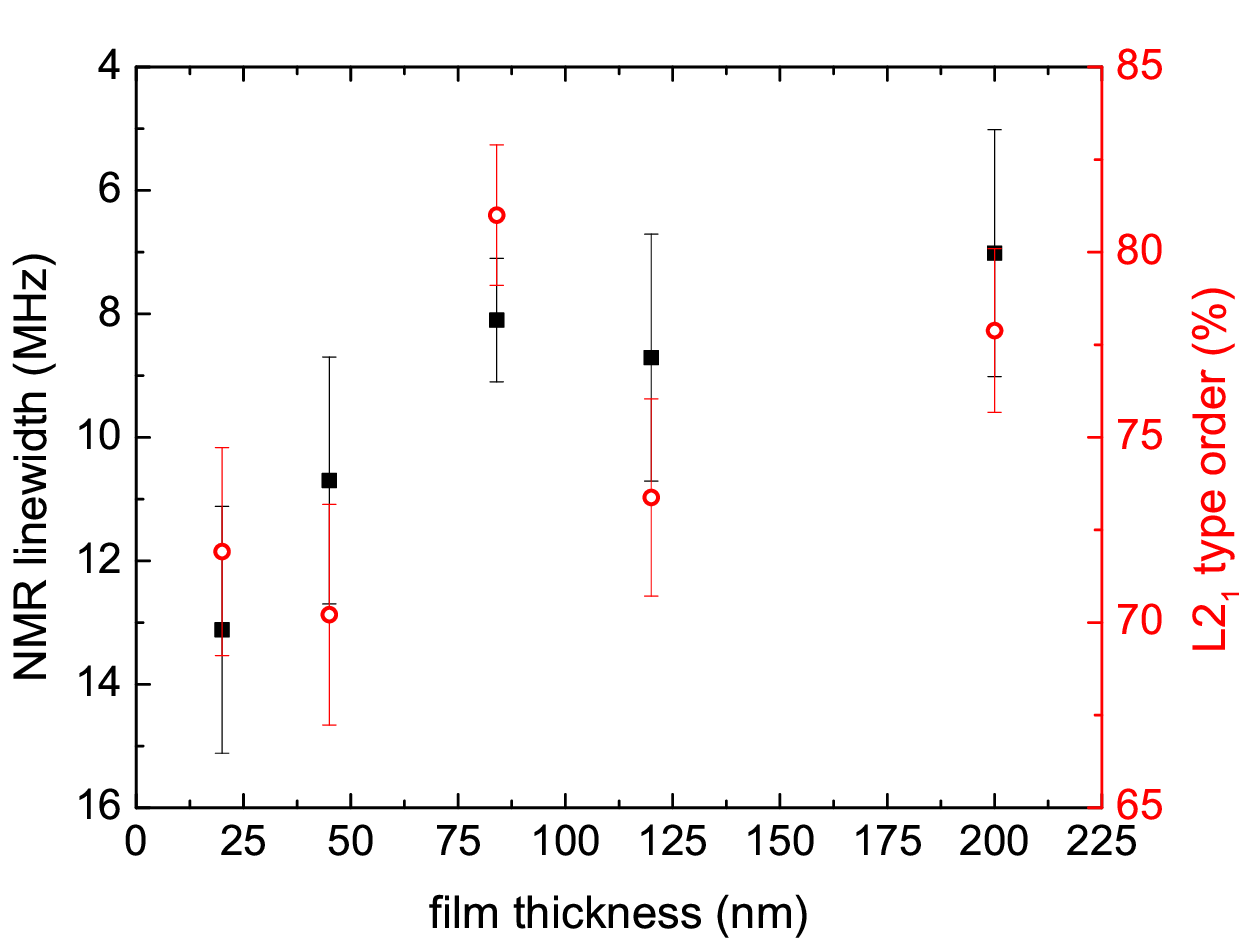}
\caption{(Color online) $L2_1$ type order contribution (red open circles) and NMR linewidth (black squares) as a function of film thickness. }
\label{Fig_4}
\end{figure}

Our results confirm a quite high degree of order for all film
thicknesses (Fig.~\ref{Fig_4}). The highest degree of $L2_1$ order
is found to be as high as 81\% for the 84~nm film. 
In order to further validate
our analysis, we compare the trend in NMR linewidth for all films
as function of thickness. 
Figure~\ref{Fig_4} shows that the linewidth
decreases with increasing film thickness, indicating an
improvement of ordering in thicker films and/or release of strain (please note that the linewidth axis is inverted to allow for a more direct comparison between evolution of linewidth and degree of order).
Generally, the dependence of the NMR linewidth reflects the evolution of $L2_1$ ordering, as expected. Interestingly, the 84~nm film sticks out demonstrating smallest linewidth along with highest degree of order of about 81\%. We will come back to the peculiarities of the 81~nm thick film at a later point.

\begin{figure}
\centering
\includegraphics[width=8cm]{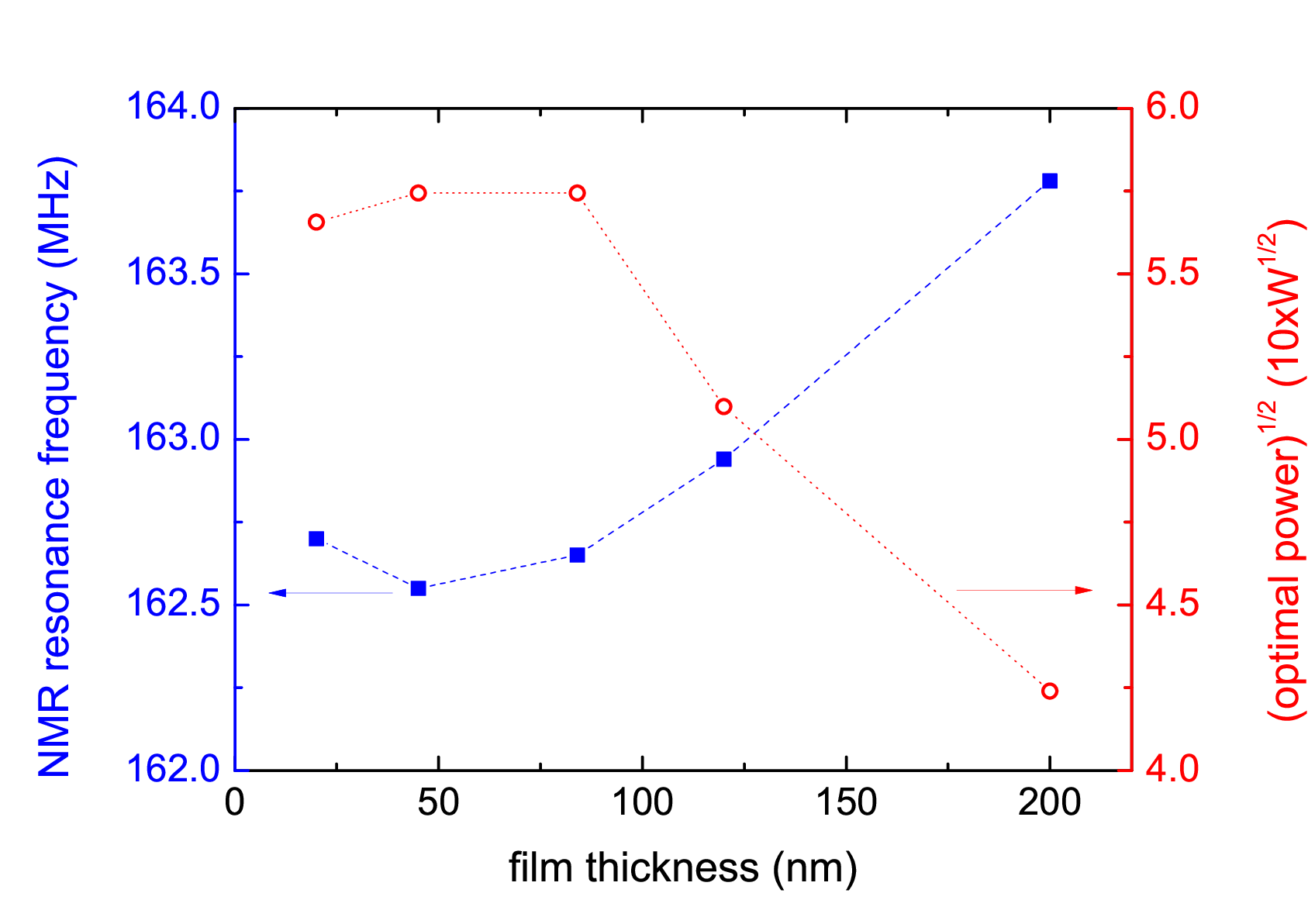}
\caption{(Color online) Thickness dependence of NMR resonance frequency (blue
squares, left side) and square root of the optimal power (red
circles, right side) for the main line ($\sim163$~MHz) of the $^{59}$Co
NMR spectrum.}
\label{Fig_2}
\end{figure}

In order to further shed light on the relation between structural quality and film thickness, we analyzed the thickness dependence of the square root of
the optimal power (figure~\ref{Fig_2}, red circles, right side) since the measurement of optimal power (i.e.\ the power producing the maximum spin-echo intensity) of the applied rf pulses allows to indirectly investigate the magnetic stiffness or magnetic anisotropy on a local scale via monitoring the restoring field;\cite{pan97,WK08} specifically, the square root of the optimal power is proportional to the restoring field. The analysis of the local restoring field is interesting for the investigation of the quality of thin films with respect to their thickness as explained in the following. Typically, there is a critical thickness for films with an epitaxial relation between film and substrate: While films below a certain thickness show uniform full strain - either tensile or compressive, depending on the ratio between the lattice constants - films above the critical thickness release strain. The critical thickness may depend on several parameters such as ratio of lattice constants between film and substrate and elastic properties of film material.\cite{WBA96} The release of strain in films with their thickness exceeding the critical value may lead to dislocations, defects and disorder accompanied by a change in magnetic anisotropy;\cite{WBA96} we may monitor this effect by measuring the restoring field by NMR. In the present case of Co$_2$FeAl$_{0.5}$Si$_{0.5}$ films, we find that the restoring field of the 200-nm film reaches the value of the bulk sample consistent with a full release of strain and negligible magnetic anisotropy consistent with a cubic system. Interestingly, there is a clear transition for the optimal power at thicknesses between 84 and 120~nm (open circles in Fig.~\ref{Fig_2}). This transition for the optimal power at thicknesses between 84 and 120~nm may be related to the fact that the Co$_2$FeAl$_{0.5}$Si$_{0.5}$ films are mostly strained at thicknesses below 100~nm and start to relax above 100~nm as observed by XRD.\cite{PAB13} In the following, let us combine the information on 
thickness dependence of the linewidth, amount of $L2_1$ order and local magnetic anisotropy to understand the evolution of film quality in films with different thickness.
The Co$_2$FeAl$_{0.5}$Si$_{0.5}$ films under ~100 nm
thick are fully strained with a tetragonal distortion and remain, hence,
structurally uniform while above 100~nm, the films start to
relax, which leads to a lower quality of thicker films. Since for Co$_2$FeAl$_{0.5}$Si$_{0.5}$, the critical thickness is about 100~nm, the best structural quality with the
highest $L2_1$ ordering is observed in the largest available
thickness below 100 nm, i.e., 84-nm Co$_2$FeAl$_{0.5}$Si$_{0.5}$
film.

In general, the resonance frequencies of the films are higher than in the bulk sample and closer to the average value (165~MHz) between highly ordered Co$_2$FeSi (139~MHz) and $B2$ ordered Co$_2$FeAl (190~MHz). Three factors may contribute to the evolution of resonance frequencies: (i) strain, hence, the frequency may scale with the film thickness, (ii) stoichiometry, and  (iii) degree of order. (ii) and (iii) are both based on the fact that the stoichiometry, in particular the Fe stoichiometry, and order are both closely linked to the magnetic moment of local neighbors and, hence, in turn to the local hyperfine field and frequency via the transferred fields.

Since our analysis of the local magnetic anisotropy revealed a relation between strain and thickness in our Co$_2$FeAl$_{0.5}$Si$_{0.5}$ films, we study the impact of strain on the NMR resonance frequencies using their evolution as function of film thickness (Fig.~\ref{Fig_2}). The resonance frequencies are more or less constant for films thinner than  84~nm and significantly increase with increasing film thickness. Interestingly, a matching inverse trend between local restoring field and film thickness or in other words a similar transition between resonance frequencies and optimal rf power at thicknesses between 84 and 120~nm is observed, confirming that the release of strain contributes to the evolution of resonance frequencies.

Besides $B2$ type ordering, we also observe Fe excess at the expense of Al and Si. We were able to
quantify this off-stoichiometry by fitting the data with
Eq.\ref{eq1}; as a results, we obtain about 8 to 13\% excess Fe in the films indicating that the film composition differs from that of the target. This may be understood as follows: In the off-axis sputtering geometry, the substrate is positioned at an angle of $\cong$55$^\circ$ with respect to the normal direction of the sputter target. This arrangement is crucial in minimizing the energetic bombardment damage of the sputtered atoms on the film. Due to this angled deposition, sometimes there is a difference in the stoichiometry of the arriving atoms at the substrate as compared to the target composition. Film compositions different from the corresponding target when using on-axis sputtering were already reported previously, as, e.g., stoichiometric Co$_2$MnSi films are obtained from stoichiometry adjusted targets.\cite{SMO06,OSN06}

\begin{figure}[b]
\centering
\includegraphics[width=8cm]{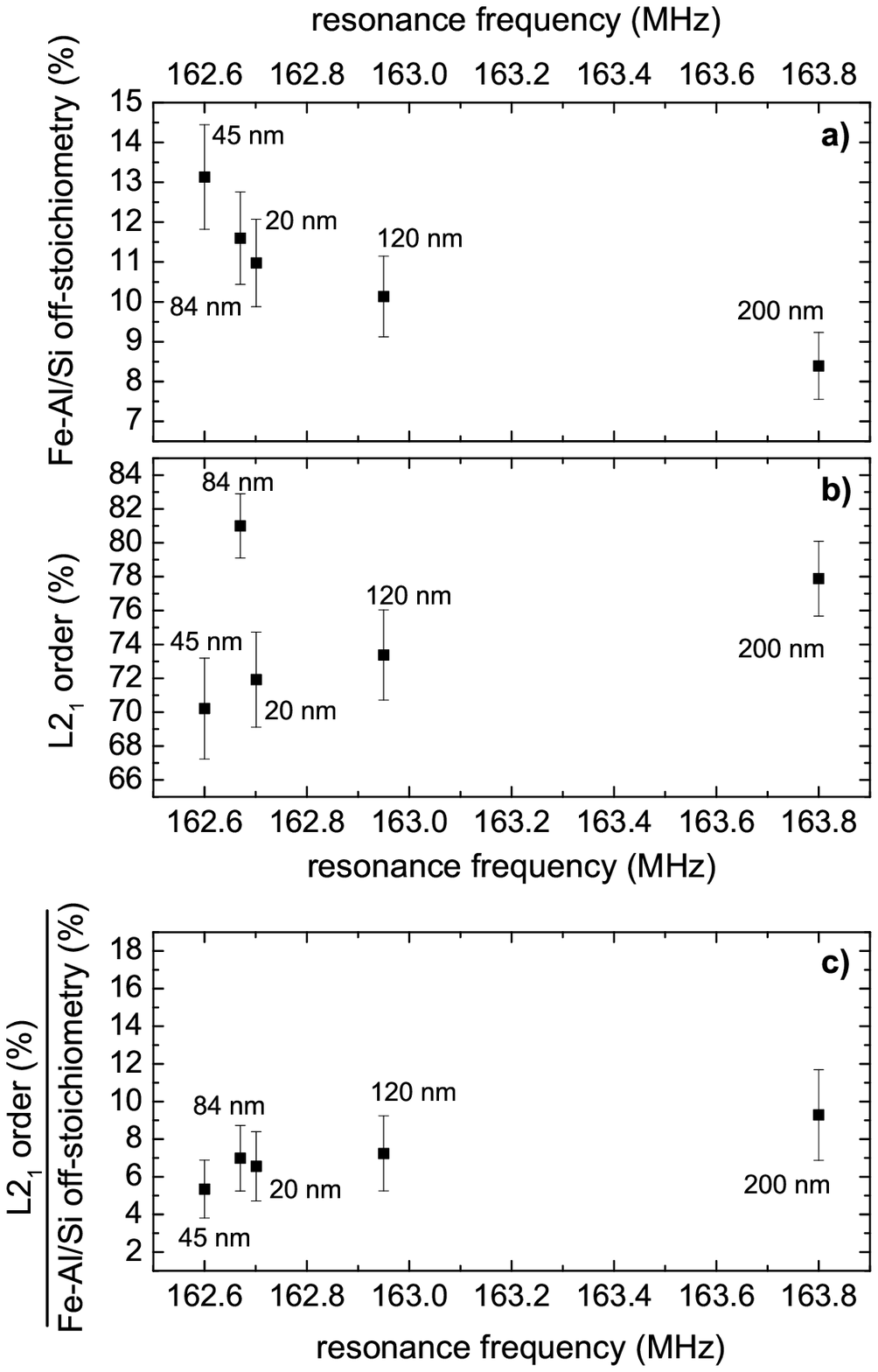}
\caption{Fe/(Al+Si) off-stoichiometry (\textbf{a}),  $L2_1$ ordering (\textbf{b}) and their ratio (\textbf{c}) of the Co$_2$FeAl$_{0.5}$Si$_{0.5}$ films as a function of resonance frequency of the main NMR line.}
\label{Fig_5}
\end{figure}

For further analysis of the contribution of Fe-stoichiometry on the resonance frequencies, we plotted the relation between the Fe to Al/Si off-stoichiometry and the corresponding resonance frequency of the main NMR line (Fig.~\ref{Fig_5}(a)). The NMR resonance frequency monotonously increases with decreasing amount of off-stoichiometry. Lower level of off-stoichiometry implies lower macroscopic magnetic moment of the sample, which in turn due to a negative hyperfine constant of Co \cite{WJS12,AAS90} yields higher resonance frequency consistent with our observations (Fig.~\ref{Fig_5}(a)).

\begin{figure}[b]
\centering
\includegraphics[width=8cm]{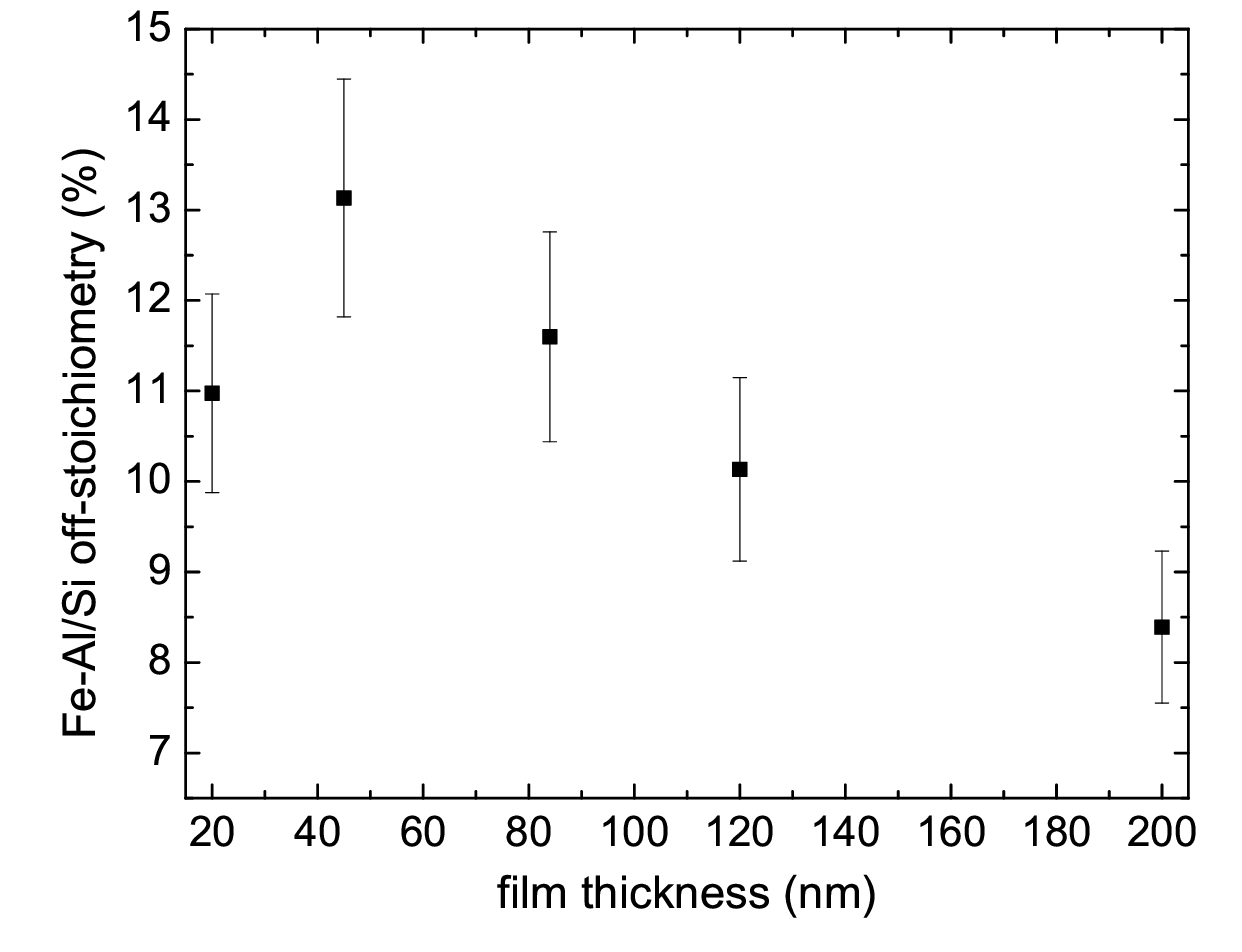}
\caption{Fe/(Al+Si) off-stoichiometry as function of film thickness.}
\label{Fig_stoivsthick}
\end{figure}

Interestingly, the Fe-stoichiometry scales also with the film thickness (see Fig.~\ref{Fig_stoivsthick}) with the thinnest film being the exception from the trend. This observation may also be related to the thick films being more similar to the bulk samples.

The binomial analysis shows much larger contributions from L2$_1$ order in the present films ($70\%-81\%$) than in the bulk sample ($59\%$, see Ref.~\onlinecite{WKS12}). Hence, the shift of the resonance frequencies towards the mean frequency value between highly ordered Co$_2$FeSi and $B2$ ordered Co$_2$FeAl may relate, at least partially, to the higher order in the films compared to the bulk. This working hypothesis is confirmed by the dependence of the L2$_1$ order as the function of resonance frequency (see Fig.~\ref{Fig_5}(b)), here, higher degree of L2$_1$ order yields a higher frequency, with the 84~nm film being the only exception. Please note, that a similar shift of the resonance line of Co$_2$FeAl$_{0.5}$Si$_{0.5}$ films in response to the degree of order has been shown by Inomata \textit{et al.} but not commented.\cite{IWJ08} Since our films have always a higher degree of $L2_1$ type order than the bulk we may conclude that the transformation to the higher ordered $L2_1$ type phase depends rather on kinetic than thermodynamic control. This competition between thermodynamic and kinetic control has a strong dependence on the film thickness since the ordering depends on the thermal history of a given sample and hence on the substrate temperature (which is constant and it is 600$^{\circ}$~C in all films), the cooling rate, but also on the defect concentration and the length of mean free path, viz. the diffusion length of atoms during the ordering process.     
	
Additionally, if we plot the ratio between L2$_1$ order and Fe to Al/Si off-stoichiometry (see Fig.~\ref{Fig_5}(c)) as the function of resonance frequency, we clearly see a monotonous shift of the NMR frequency towards the mean value between Co$_2$FeSi and Co$_2$FeAl, which further proves the scenario of the interplay between strain, stoichiometry and structural order.

\section{Summary}

We presented a detailed NMR analysis of the structural and local
magnetic properties of Co$_2$FeAl$_{0.5}$Si$_{0.5}$ films with
varying thickness. Our findings are classified in two categories. (i) We confirm a new growth technique to yield Heusler films with an high quality, which may open way to enhance the performance of Heusler compounds in spintronic devices in general, if this technique is established. We prove the quality of the films by detailed NMR analysis of the film properties. (ii) We use the NMR technique to disentangle different contributions to the film quality, namely film thickness, its impact on strain and local anisotropy, stoichiometry and degree of order.

\section{Acknowledgments}
This work is supported by the Materials World Network grant from
the National Science Foundation (DMR-1107637) and from Deutsche
Forschungsgemeinschaft DFG (WU595/5-1). Partial support is
provided by the NanoSystems Laboratory at the Ohio State
University. SW gratefully acknowledges funding by DFG in the Emmy
Noether programme, project WU595/3-1. We thank P. Woodward and W. L\"oser for
discussion and C.G.F. Blum for help with the preparation of the
sputter targets.

\pagebreak



\section{Supplemental material}

\subsection*{Microstructure and binomial model}

The coefficient $C_{B2}$ and
$C_{L2_1}$ represent the relative contribution from $^{59}$Co
nuclei with a $B2$ and $L2_1$ first shell environment,
respectively, and $C_{B2}$ + $C_{L2_1}$ = 1. However, this calculational model does not allow any statements on how those phases manifest in the microstructure. There are two borderline cases of how to realize the presence of both $L2_1$ and $B2$ in a given sample. Case (i) deals with large $B2$ type domains within a $L2_1$ matrix, where the number of Co nuclei located at the interfaces between both phases are small and negligible compared to the number of Co nuclei within a certain phase region in line with the recent report on Co$_2$MnSi films by  T. Miyajima \textit{et al.}\cite{MOK09} In that case, the coefficients as obtained from our binomial model immediately give the ratio between $L2_1$ and $B2$ phases. 
In the second borderline case (ii) both $L2_1$ and $B2$ phases are very finely dispersed, in the most extreme case each $B2$ ordered unit cell has 26 neighboring unit cells at the interface to an $L2_1$-ordered cell. Since the Co nuclei in those interface cells would also have the same environments with different numbers of Fe and Si,Al atoms as in a $B2$ type cell one would expect the very same hyperfine fields as for those which are in the center of a large $B2$ cluster; however, the distribution of atoms would not follow the binomial distribution function as given in Eq.~1, in particular, the new distribution function would be also symmetric around the main line. In general, the $B2$ contributions represented by the $C_{B2}$ coefficient in our model in the case of scenario (ii) represent the number of Co nuclei in an interface and in a $B2$ cell, hence, the overall degree of order is even underestimated if a finely dispersed microstructure is present. Since our binomial model as given in Eq.1 nicely describes our NMR data, scenario (i) seems to be valid in the present case in line with the recent report on Co$_2$MnSi films by T. Miyajima \textit{et al.}\cite{MOK09}

\subsection*{Distribution of Al and Si}

Compounds with Si have extra valence electron with respect to the compounds with Al. This extra valence electron increases the magnetic moment, as also evident in the magnetization data (see e.g. Ref.~\onlinecite{BFF07,UOK12}). Since the transferred contributions to the hyperfine field depends on the number and magnetic moments of neighboring atoms, the hyperfine field of a given Co nucleus will be different depending on whether a Si or an Al atom is in the first coordination shell. Hence, each specific configuration with particular Al and Si neighbors in the first shell of Co, arising from the random distribution expected for a quaternary compound, will have a different resonance frequency. Since the difference if only Al or Si are intermixed will be small, the difference will be more pronounced if also Fe is intermixed. For details see Ref.~\onlinecite{WKS12}

\begin{figure}[htb]
\centering
\includegraphics[width=8cm]{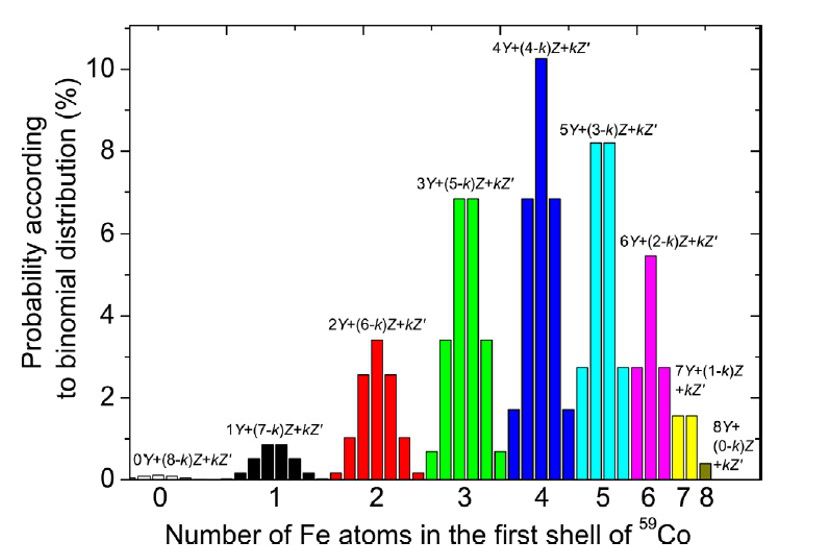}
\caption{(Color online) The relative probabilities calculated using Eq.2 in Ref.~\onlinecite{WKS12} corresponding to a $B2$ structure with a random distribution of the Al and Si atoms. Note, that lines corresponding to the same number of Fe atoms have the same color demonstrating that the sub-lines are related to the random distribution of Al and Si (picture taken from Ref.~\onlinecite{WKS12}). }
\label{Fig_supp}
\end{figure}

Assuming a $B2$ type structure with an additional random distribution of Al and Si, the
probability $P(n, k, m, x, y)$ for a particular surrounding of the form \textit{n Fe + (n - k) Al + k Si}
atoms in a certain shell of the Co atom is now given by a multinomial distribution, which
depends on the ratio of the intermixing between the atoms on
the 1$a$ position, and on the ratio of the partial substitution of
Al by Si as described in our manuscript. The result of this model which calculates the probabilities (neglecting the contributions from Fe off-stoichiometry) is shown in Fig.~\ref{Fig_supp}. In order to demonstrate the
effects of the random distribution of Al and Si, lines corresponding to the same number of Fe
atoms have the same color. The main lines with the same number of Fe atoms in the first
shell have two sub-lines with the maximum intensity for an odd number of Fe atoms and one sub-line with the maximum intensity for an even number of Fe atoms. This leads also to the observation that the main
lines corresponding to an odd number of $Y$ atoms are broader and that the probability of the
corresponding most prominent sub-line is lower. The
sum of the areas of, i.e., the \textit{5 Fe +(3- k)Al+ k Si} main line are
the same as for the corresponding \textit{3 Fe +(5- k)Al+ k Si} main
lines. This follows from the symmetry of the binomial distribution
with $x$, $y$=0.5. Moreover, the number of sub-lines
decreases with increasing resonance frequency of the main
line.\cite{WKS12} This intrinsic asymmetry of the main lines arising from the Fe, Al, Si requires that the distribution of Al and Si is taken and, let us stress here, need to be taken into account in order to rate the correct Fe-stoichiometry.

\begin{thebibliography}{10}


\bibitem{GME83}
R.~A. de~Groot, F.~M. M\"uller, P.~G.~van Engen, and K.~H.~J. Buschow, Phys. Rev. Lett. \textbf{50}, 2024 (1983).

\bibitem{CVB02}
J.~M.~D. Coey, M.~Venkatesan, and M.~A. Bari, in \emph{Lecture Notes in Physics}, edited by C.~Berthier, L.~P. Levy, and G.~Martinez (Springer-Verlag, Heidelberg, 2002), Vol. 595, p. 377 -- 396, \emph{"Half-Metallic Ferromagnets"}.

\bibitem{FFB07}
C.~Felser, G.~H. Fecher, and B.~Balke, Angewandte, Intern. Ed. \textbf{46}, 668 (2007).

\bibitem{IIT08}
K.~Inomata, N.~Ikeda, N.~Tezuka, R.~Goto, S.~Sugimoto, M.~Wojcik, and
  E.~Jedryka, Sci. Technol. Adv. Mater. \textbf{9}, 014101 (2008).

\bibitem{GFP11}
T.~Graf, C.~Felser, and S.~S.~P. Parkin, Progress in Solid State Chemistry \textbf{39}, 1--50 (2011).

\bibitem{BBV13}
D.~Bombor, C.~G.~F. Blum, O.~Volkonskiy, S.~Rodan, S.~Wurmehl, C.~Hess, and
  B.~B\"uchner, Phys. Rev. Lett. \textbf{110}, 066601 (2013).

\bibitem{WFK05b}
S.~Wurmehl, G.~H. Fecher, H.~C. Kandpal, V.~Ksenofontov, C.~Felser, H.-J. Lin, and J.~Morais, Phys. Rev. B \textbf{72}, 184434 (2005).

\bibitem{WFK06}
S.~Wurmehl, G.~H. Fecher, H.~C. Kandpal, V.~Ksenofontov, C.~Felser, and H.-J. Lin, Appl. Phys. Lett. \textbf{88}, 032503 (2006).

\bibitem{TTO09}
T.~Kubota, S.~Tsunegi, M.~Oogane, S.~Mizukami, T.~Miyazaki, H.~Naganuma, and Y.~Ando, Appl. Phys. Lett. \textbf{94}, 122504 (2009).

\bibitem{IWJ08}
K.~Inomata, M.~Wojcik, E.~Jedryka, N.~Ikeda, and N.~Tezuka, Phys. Rev. B \textbf{77}, 214425 (2008).

\bibitem{pan97}
P.~Panissod, in \emph{NATO ASI series High Tech}, edited by V. G. Baryakhtar, P. E. Wigen, and N. A. Lesnik (Kluwer Academic, Dordrecht, 1997), Vol. 48, p. 225, \emph{"Structural and Magnetic Investigations of Ferromagnets by NMR. Application to Magnetic Metallic Multilayers"}.

\bibitem{IOM06}
K.~Inomata, S.~Okamura, A.~Miyazaki, N.~Tezuka, M.~W\'ojcik, and E.~Jedryka, J. Phys. D: Appl. Phys. \textbf{39}, 816 (2006).

\bibitem{WK08}
S.~Wurmehl and J.~T. Kohlhepp, Journal of Physics D: Applied Physics \textbf{41}, 173002 (2008).

\bibitem{WKS07}
S.~Wurmehl, J.~T. Kohlhepp, H.~J.~M. Swagten, B.~Koopmans, M.~Wojcik, B.~Balke, C.~G.~H. Blum, V.~Ksenofontov, G.~H. Fecher, and C.~Felser, Appl. Phys. Lett. \textbf{91}, 052506 (2007).

\bibitem{WKS08}
S.~Wurmehl, J.~T. Kohlhepp, H.~J.~M. Swagten, and B.~Koopmans, J. Phys. D: Appl. Phys. \textbf{41}, 115007 (2008).

\bibitem{WKS09}
S.~Wurmehl, J.~T. Kohlhepp, H.~J.~M. Swagten, B.~Koopmans, C.~G.~F. Blum, V.~Ksenofontov, H.~Schneider, G.~Jakob, D.~Ebke, and G.~Reiss, J. Phys. D: Appl. Phys. \textbf{42}, 084017 (2009).

\bibitem{WJK11}
S.~Wurmehl, P.J. Jacobs, J.T. Kohlhepp, H.J.M. Swagten, B.~Koopmans, M.J.~Carey, S.~Maat, and J.R. Childress, Appl. Phys. Lett. \textbf{98}, 12506 (2011).

\bibitem{RAB13}
S.~Rodan, A.~Alfonsov, M.~Belesi, F.~Ferraro, J.~T. Kohlhepp, H.~J.~M. Swagten, B.~Koopmans, Y.~Sakuraba, S.~Bosu, K.~Takanshi, B.~B\"uchner, and S.~Wurmehl, Appl. Phys. Lett. \textbf{102}, 242404 (2013).

\bibitem{BWF07}
B.~Balke, S.~Wurmehl, G.~H. Fecher, C.~Felser, Maria C.~M. Alves, F.~Bernardi, and J.~Morais, Appl. Phys. Lett. \textbf{90}, 172501 (2007).

\bibitem{BFF07}
B.~Balke, G.~H. Fecher, and C.~Felser, Appl. Phys. Lett. \textbf{90}, 242503 (2007).

\bibitem{WJS12}
M.~W\'ojcik, E.~Jedryka, H.~Sukegawa, T.~Nakatani, and K.~Inomata, Phys. Rev. B \textbf{85}, 100401 (2012).

\bibitem{TIM06}
N.~Tezuka, N.~Ikeda, F.~Mitsuhashi, and S.~Sugimoto, Appl. Phys. Lett. \textbf{89}, 112514 (2006).

\bibitem{TIS06}
N.~Tezuka, N.~Ikeda, S.~Sugimoto, and K.~Inomata, Appl. Phys. Lett. \textbf{89}, 252508 (2006).

\bibitem{TIM07}
N.~Tezuka, N.~Ikeda, F.~Mitsuhashi, A.~Miyazaki, S.~Okamura, M.~Kikuchi, S.~Sugimoto, and K.~Inomata, J. Magn. Magn. Mater. \textbf{310}, 1940 (2007).

\bibitem{TIS07b}
N.~Tezuka, N.~Ikeda, S.~Sugimoto, and K.~Inomata, J. Appl. Phys. \textbf{46}, L454 (2007).

\bibitem{TIM09}
N.~Tezuka, N.~Ikeda, F.~Mitsuhashi, and S.~Sugimoto, Appl. Phys. Lett. \textbf{94}, 162504 (2009).

\bibitem{NRG07}
T.~M. Nakatani, A.~Rajanikanth, Z.~Gercsi, Y.~K. Takahasi, K.~Inomata, and K.~Hono, J. Appl. Phys. \textbf{102}, 033916 (2007).

\bibitem{SSW09}
R.~Shan, H.~Sukegawa, W.~H. Wang, M.~Kodzuka, T.~Furubayashi, T.~Ohkubo, S.~Mitani, K.~Inomata, and K.~Hono, Phys. Rev. Lett. \textbf{102}, 246601 (2009).

\bibitem{TMS12}
N.~Tezuka, F.~Mitsuhashi, and S.~Sugimoto, J. Appl. Phys. \textbf{111}, 07C718 (2012).

\bibitem{Tez12}
N.~Tezuka, J. Magn. Magn. Mater. \textbf{324}, 3588 (2012).

\bibitem{PAB13}
B.~Peters, A.~Alfonsov, C.~G.~F. Blum, P.~M. Woodward, S.~Wurmehl, B.~B\"uchner, and F.~Y. Yang, Appl. Phys. Lett. \textbf{103}, 162404 (2013).

\bibitem{WKS12}
S.~Wurmehl, J.~T. Kohlhepp, H.~J.~M. Swagten, and B.~Koopmans, J. Appl. Phys \textbf{111}, 043903 (2012).

\bibitem{Koz99}
T.~Kozakai, R.~Okamoto, and T.~Miyazaki, Zeitschrift f\"ur Metallkunde \textbf{90}, 261 (1999).

\bibitem{UOK12}
R. Y. Umetsu, A. Okubo, and R. Kainuma, J. Appl. Phys. \textbf{111}, 073909 (2012).

\bibitem{MOK09}  
T. Miyajima, M. Oogane, Y. Kotaka, T. Yamazaki, M. Tsukada, Y. Kataoka, H. Naganuma, and Y. Ando, Appl. Phys. Express \textbf{2}, 093001 (2009).

\bibitem{WAK13}
S.~Wurmehl, A.~Alfonsov, J.~T. Kohlhepp, H.~J.~M. Swagten, B.~Koopmans, M.~W\'ojcik, B.~Balke,  V.~Ksenofontov, C.~G.~H. Blum, and B.~B\"uchner, Phys. Rev. B \textbf{88}, 134424 (2013).

\bibitem{WBA96}
W.~Weber, A.~Bischof, R.~Allenspach, C.~H. Back, J.~Fassbender, U.~May, B.~Schirmer, R.~M. Jungblut, G.~G\"untherodt, and B.~Hillebrands, Phys. Rev. B \textbf{54}, 4075 (1996).

\bibitem{SMO06}
Y.~Sakuraba, T.~Miyakoshi, M.~Oogane, Y.~Ando, A.~Sakuma, T.~Miyazaki, and H.~Kubota, Appl. Phys. Lett. \textbf{89}, 052508 (2006).

\bibitem{OSN06}
M.~Oogane, Y.~Sakuraba, J.~Nakata, H.~Kubota, Y.~Ando, A.~Sakuma, and T.~Miyazaki, J. Phys. D: Appl. Phys. \textbf{39}, 834 (2006).

\bibitem{AAS90}
H.~Akai, M.~Akai, S.~Bl\"ugel, B.~Drittler, H.~Ebert, K.~Terakura, R.~Zeller, and P.~H. Dederichs, Prog. Theor. Phys. Suppl. \textbf{101}, 11 (1990).




\end{thebibliography}
\end{document}